\documentclass{mn2e}
\usepackage{graphicx}
\usepackage{verbatim}
\usepackage{amsmath, amssymb}

\newcommand{\arcm}{{$^\prime\,$}}
\newcommand{\arcs}{{$^{\prime\prime}\,$}}    
\newcommand{\mnras}{{\it MNRAS}}

\newcommand{\apj}{{\it ApJ}}
\newcommand{\apjl}{{\it ApJL}}
\newcommand{\apjs}{{\it ApJS}}
\newcommand{\procspie}{{\it Proc. SPIE}}

\begin{document}

\title{Shedding Light on the Matter of Abell 781}

\author[D. Wittman et al.]{D. Wittman\thanks{E-mail: dwittman@physics.ucdavis.edu}, William Dawson, and Bryant Benson\\
Physics Department, University of California, Davis,  CA 95616}
\maketitle

\begin{abstract} The galaxy cluster Abell 781 West has been viewed as
  a challenge to weak gravitational lensing mass calibration, as Cook
  and dell'Antonio (2012) found that the weak lensing signal-to-noise
  in three independent sets of observations was consistently lower
  than expected from mass models based on X-ray and dynamical
  measurements.  We correct some errors in statistical inference in
  Cook and dell'Antonio (2012) and show that their own results agree
  well with the dynamical mass and exhibit {\it at most}
  2.2--2.9$\sigma$ low compared to the X-ray mass, similar to the tension
  between the dynamical and X-ray masses.  Replacing their simple
  magnitude cut with weights based on source photometric redshifts
  eliminates the tension between lensing and X-ray masses; in this
  case the weak lensing mass estimate is actually higher than, but
  still in agreement with, the dynamical estimate.  A comparison of
  lensing analyses with and without photometric redshifts shows that a
  1--2$\sigma$ chance alignment of low-redshift sources lowers the
  signal-to-noise observed by all previous studies which used
  magnitude cuts rather than photometric redshifts.  The fluctuation
  is unexceptional, but appeared to be highly significant in Cook and
  dell'Antonio (2012) due to the errors in statistical interpretation.
\end{abstract}

\begin{keywords}gravitational lensing: weak--methods: statistical--galaxies: clusters: individual: Abell 781\end{keywords}

\section{Introduction}\label{sec-intro}

Abell 781 is the collective name for four clusters of galaxies, not
all of which are physically related. Cook and dell'Antonio (2012,
hereafter C12) argue that weak lensing mass estimates of one of these
clusters are substantially lower than expected based on X-ray and
dynamical estimates, and that the persistence of this effect across
multiple data sets from multiple telescopes presents a challenge to
weak lensing calibration.  In this paper we show that this claim does
not stand up to closer scrutiny, and we attempt to examine the issues
in a way which sheds light on weak lensing analyses in general.  In
this introduction we summarize previous studies to help the reader
understand more specifically what the claim is, and then outline our
response.

\begin{figure}
\centerline{\includegraphics[scale=0.75]{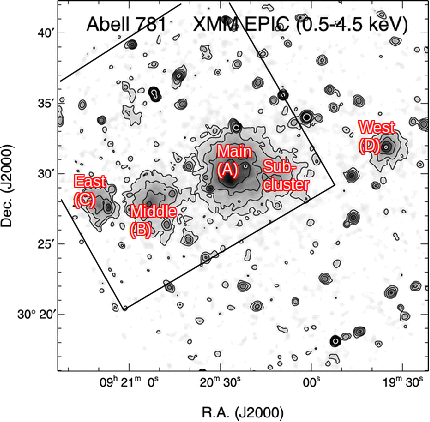} }
\caption{XMM image of the Abell 781 region, with Sehgal {\it et al.}
  (2008) labels (words) and Cook \& dell'Antonio (2012) labels
  (letters).  The smaller square indicates the {\it Chandra} field of
  view.  Adapted from Sehgal {\it et al.} (2008).
 \label{fig-xmm}}
\end{figure}

Figure~\ref{fig-xmm} is a map of the region as seen by XMM.  In this
figure, adapted from Sehgal {\it et al.} (2008, hereafter S08), we
have labeled the four components (East, Middle, Main, and West) as
named by those authors and as labeled by C12 (C, B, A, and D
respectively).  The Main cluster ($z=0.3004$, Geller {\it et al.}
2010) is the brightest in both galaxies and X-rays and is close to the
original Abell, Corwin \& Olowin (1989) position, and is therefore
most strongly identified with the Abell 781 label.  This cluster
appears to be in a merging state based on the offset Subcluster gas
distribution labeled in Figure~\ref{fig-xmm}.  The Middle cluster (so
called because it is the middle of the three seen in archival {\it
  Chandra} data also studied by S08) is about 6\arcm\ to the east of
the Main cluster and is at the same redshift, within a few thousand
kilometers per second ($z=0.2915$, Geller {\it et al.}  2010).  The
East cluster is an additional $\sim 3$\arcm\ further east, but is at
$z=0.4265$ (Geller {\it et al.}  2010) and thus physically unrelated
to Main and Middle.  The West cluster, projected $\sim$11\arcm\ west
of Main, is at $z=0.4273$ (Geller {\it et al.}  2010) and thus is
physically near East, with a transverse separation of
$\sim$21\arcm\ or 7 Mpc.


Abell 781 was identified by Wittman {\it et al.} (2006) as a
shear-selected cluster in the Deep Lens Survey (DLS; Wittman {\it et
  al.} 2002), and Abate {\it et al.} (2009) performed weak lensing
analyses of extended X-ray sources identified in {\it Chandra}
followup of those shear-selected clusters.  They fit
Navarro-Frenk-White (NFW; Navarro, Frenk \& White 1997) profiles to
Main, Middle, East, and Subcluster, which they called a, b, c, and d
respectively.  They did not fit the West cluster because, being
outside the {\it Chandra} field of view, it did not meet their
selection criteria.  Because ``d'' in Abate {\it et al.}  (2009) and
``D'' in C12 refer to different structures, we will use the S08
nomenclature to avoid confusion.

%

\begin{figure}
\centerline{\includegraphics[scale=0.52]{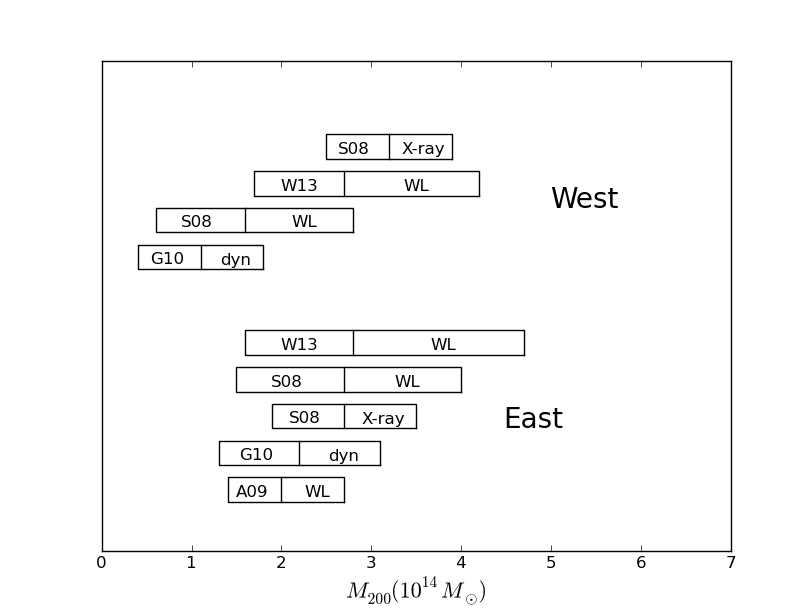} }
\caption{A graphical summary of the mass estimates for East and West
  (also listed in Table~1) clearly shows that all published weak
  lensing mass estimates are consistent with both X-ray and dynamical
  values. The dynamical mass estimate of West is actually the one in
  greatest tension with the X-ray mass estimate.  The vertical line
  segments in each bar indicate the best-fit value and the $\pm
  1\sigma$ uncertainties.  Paper abbreviations are: S08, Sehgal {\it
    et al.} (2008); A09, Abate {\it et al.} (2009); G10, Geller {\it
    et al.}  2010); and W13, this work.  Some of the estimates have
  been converted to $M_{200}$ as noted in Table~1.
 \label{fig-massest}}
\end{figure}

C12 were motivated to investigate West in more detail because in DLS
convergence maps (Wittman {\it et al.} 2006, Kubo {\it et al.}  2009,
Khiabanian \& Dell’Antonio 2008) West appears at lower significance
than East despite their consistent (within 1$\sigma$) X-ray
temperatures and X-ray inferred masses (S08; see
Table~\ref{tab-massestimates} and Figure~\ref{fig-massest} for a
summary of all relevant mass estimates).  However, C12 do not mention
that S08 also fit NFW profiles to the DLS lensing data and found
$M_{200}=2.7^{+1.3}_{-1.2} \times 10^{14} M_\odot$ and
$1.6^{+1.2}_{-1.0} \times 10^{14} M_\odot$ for East and West
respectively\footnote{These numbers have been converted from the
  $M_{500}$ estimates of S08 to facilitate comparison with the
  $M_{200}$ estimates given by C12.}; {\it i.e.}, the difference is
within 1$\sigma$.  Therefore, if the less prominent appearance of West
on the previously published convergence maps is significant, the
problem may lie in the construction or interpretation of these
convergence maps rather than weak lensing generally.

\begin{table*}
\begin{minipage}{126mm}
\caption{Abell 781 Mass Estimates (M$_{200}$ in units of $10^{14}$\,M$_\odot$).  
The S08 values have been converted from $M_{500}$, and the Geller et  al. (2010)
values have been converted from velocity dispersion by C12.}
\begin{tabular}{lccccc}
 & S08 & S08 & Abate et al. (2009) & Geller et  al. (2010) & This  Work\\
Cluster & X-ray & Weak lensing & Weak lensing & Galaxy velocities & Weak lensing\\ \hline
{\bf East}  &  {\bf 2.7$\pm$0.8}  & {\bf 2.7$^{+1.3}_{-1.2}$} &  {\bf
    2.0$^{+0.7}_{-0.6}$}  &  {\bf 2.2$\pm$0.9}  &  {\bf  2.8$^{+1.9}_{-1.2}$} \\
{\bf  West} &  {\bf 3.2$\pm$0.7} & {\bf  1.6$^{+1.2}_{-1.0}$}  &
(no data)  &  {\bf 1.1$\pm$0.7}  &  {\bf 2.7$^{+1.5}_{-1.0}$} \\ \hline
Main 	&  8.0$\pm$1.1  & 4.0$^{+1.5}_{-1.3}$ & 3.5$^{+0.4}_{-0.6}$  &  4.5$\pm$3.5  &  6.7$^{+1.4}_{-1.3}$ \\
Middle  &  2.4$\pm$0.6  & 3.0$^{+1.5}_{-1.0}$ & 3.0$^{+0.6}_{-0.4}$  &  3.5$\pm$3.0  &  4.3$^{+1.6}_{-1.2}$ \\ \hline
\end{tabular}
\label{tab-massestimates}
\end{minipage}
\end{table*}

As further motivation for investigating the (perceived) low weak
lensing signal from West, C12 cite dynamical measurements which
buttress the case for roughly consistent masses in East and West:
Geller {\it et al.} 2010 find $\sigma_v=754\pm92$ and $596\pm107$ km
s$^{-1}$ for East and West respectively.  C12 converted these velocity
dispersions to $M_{200} = 2.2\pm0.9$ and $1.1\pm 0.7 \times 10^{14}
M_\odot$ for East and West respectively, and converted S08's
X-ray-derived $M_{500}$ values to $M_{200} = 2.7\pm0.8$ and $3.2\pm0.7
\times 10^{14} M_\odot$ for East and West respectively.  With all
masses in terms of $M_{200}$ it is clear that the East dynamical mass
is in excellent agreement with its X-ray mass, whereas the West
dynamical mass is 2.2$\sigma$ low compared to its X-ray mass.  We
point this out because we will show that C12's own lensing results,
properly intepreted (\S\ref{sec-stat}), are also about 2.2$\sigma$ low
compared to the X-ray model.  Whether one chooses to characterize a
2.2$\sigma$ deviation as rough agreement or as mild tension, the same
term should apply to both the dynamical-X-ray comparison and the
lensing-X-ray comparison.  Also, C12 failed to note that the S08
lensing mass estimate for West is actually {\it higher} than their
dynamical mass, further weakening the argument that weak lensing mass
estimates of West are persistently low.

The C12 investigation used three different weak lensing data sets from
three different telescopes and cameras: the original DLS imaging, new
imaging from the OPTIC camera at the WIYN telescope, and archival
SuprimeCam imaging at the Subaru telescope.  The OPTIC and SuprimeCam
data do not cover East so C12 focused on modeling West.  Rather than
estimate the mass directly, they made convergence maps and then
signal-to-noise (S/N) maps using bootstrap resampling of each data
set.  Finding relatively low S/N (1.7, 0.63, and 0.69 for DLS, OPTIC,
and SuprimeCam data respectively) at the position of the West cluster,
they asked whether this low weak lensing S/N is consistent with the
X-ray and dynamical estimates of its mass. To answer this question,
they created mock weak-lensing data sets with the mass of the West
cluster constrained by the X-ray analysis.  For an NFW halo with
concentration parameter $c=5$, they found that only 1.4\% of DLS-like
mocks yielded S/N as low as observed in their analysis of the DLS
data; only 0.2\% of OPTIC-like mocks yielded S/N as low as observed in
their analysis of the OPTIC data; and only 1.4\% of SuprimeCam-like
mocks yielded S/N as low as observed in their analysis of the
SuprimeCam data.  They also performed a similar exercise but with the
mass of the West cluster constrained by the dynamical analysis of
Geller {\it et al.}  (2010), and found somewhat better agreement, with
$\sim$10--25\% of mock realizations agreeing with any given data
set. C12 concluded that the weak-lensing signal from the West cluster
is anomalously low, that faulty point-spread function (PSF) correction
cannot be to blame because the anomaly is observed in three
independent lensing data sets, and that this anomaly provides a
``challenging obstacle'' for calibration of weak-lensing masses.

C12 also studied a strongly lensed arc at a radius of
$\sim$7\arcs\ around the West cluster.  Although the source redshift
remains unknown, C12 considered a range of likely source redshifts and
concluded that the mass contained within this radius is likely to be in
the range 1--1.5$\times$10$^{14}$ M$_\odot$.  We do not use strong
lensing information in this paper because our primary concern is the
consistency of X-ray, dynamical, and weak lensing measurements, and
because strong lensing cannot rule out any model without a secure
source redshift.

In summary, a variety of physical probes by multiple investigators
(X-ray and weak lensing by S08, and dynamical by Geller {\it et al.}
2010) support an $M_{200}$ in the range of 1-3 $\times 10^{14}
M_\odot$.  C12 do not actually infer a weak-lensing mass for the West
cluster or challenge the previously published mass estimates, but
find that their S/N for West is anomalously low given the models
inferred from X-ray and dynamical data.  Their S/N study of West was
motivated by West appearing less prominently than East in previously
published convergence maps, but they do not study East in their paper.
Therefore, we will consider two related but distinct questions: why
C12 might measure a lower S/N for West than predicted by X-ray and
dynamical models, and why West might appear less prominently than East
in previous convergence maps.


In \S\ref{sec-stat} we answer the first question by correcting some
statistical inference errors by C12 and showing that their own
modeling is entirely consistent with the dynamical model and only
modestly in tension with the X-ray model.  In \S\ref{sec-willsmap}
present a new convergence map with the DLS data weighted using source
redshift information.  The roughly equal appearance of West and East
in this map suggests an answer to the second question: previous
convergence maps suffered from a modest (1--2$\sigma$) fluctuation
in shape noise involving low-redshift sources, which is therefore
suppressed when properly deweighting low-redshift sources.  In
\S\ref{sec-summary} we summarize and discuss the implications.  In
Appendix A we investigate several factors which complicate the
interpretation of convergence maps, particularly S/N maps.  These
factors do not appear to be responsible for a substantial part of the
C12 result, but may be of interest to students of weak lensing.

\section{Statistical significance of the C12 result}\label{sec-stat}


In this section we correct some errors C12 made in statistical
inference and show that, according to their own modeling, the
significance of their result is much lower than they imply.
Throughout, we will refer to numbers from their Table~5.

First, C12 incorrectly multiplied the p-values from the different
lensing data sets to obtain an overall p-value. A correct way to
combine independent p-values is with Fisher's (1925) combined
probability test.  Given $n$ independent p-values $p_i$ where
$i=1,2,...n$, a $\chi^2$ distribution with $2n$ degrees of freedom is
obtained by summing the natural logarithms of the p-values and
multiplying by 2:
$$\chi^2 = -2\sum\limits_{i=1}^n \ln p_i$$ This $\chi^2$ value may
then be converted to a probability using standard $\chi^2$
calculators.\footnote{If multiplication of p-values seems intuitively
  correct, consider that we expect p-values of order 0.5 for correct
  models.  Multiplying p-values then yields a number of order $0.5^n$
  after $n$ independent tests.  This must approach zero for large $n$
  even if the model is correct.  Therefore multiplication cannot yield
  the correct overall p-value.}  Performing this test on the dynamical p-values
given in C12 Table~5, we find a combined p-value of 7.4\% rather than
0.3\%; in Gaussian terms, a 1.4$\sigma$ discrepancy.  For the X-ray
model in the same table, we find 0.0049\% rather than $3.92\times
10^{-5}\%$.  In other words, correcting this error alone releases the
tension between the lensing data and the dynamical model but not the
X-ray model.

Second, the lensing data sets are not actually independent.  C12 focus
on the independence of the PSF corrections, but the principal source
of uncertainty---shape noise, or the random pre-lensing orientations
of source galaxies---is not independent across the three data sets.
C12 did recognize that ``the three analyses share many of the same
galaxies'' but this point deserves much more consideration.  Shape
noise is the dominant source of random error in weak lensing.  The
three weak lensing analyses in C12 use highly overlapping sets of
sources due to similar magnitude and size cuts.  Therefore, if a shape
noise fluctuation appears in one lensing data set, it must appear with
similar strength in the others.  (An exception would be images taken
at very different wavelengths, such as visible and radio, such that
the pre-lensing shapes of galaxies are not highly correlated.)  In
other words, this dominant source of noise is highly correlated across
the three data sets.  The bootstrap resampling in C12 also drew from
the allowed range of X-ray (or dynamical) models, and this source of
variation is also correlated across the three data sets. As a
hypothetical example, if {\it each} lensing data set requires a model
from the lowest 2\% of models allowed by the X-ray fit, we should
infer that the lensing data {\it collectively} require a model from
the lowest 2\% of X-ray models rather than the lowest 0.07\% (the
result of Fisher's method assuming independent p-values), and certainly not
the lowest 0.0008\% (the result of multiplying the p-values).

Because the dominant sources of variation are highly correlated, the
most reasonable overall lensing p-value to quote is that of the
largest and most inclusive data set, the DLS.  For the X-ray model,
this value is 0.014.  In Gaussian terms, this is a 2.2$\sigma$ result
rather than a 4.9$\sigma$ result as one might infer from the ``joint
constraint'' column of C12's Table~5.  This rises to 2.9$\sigma$ if we
are concerned with the OPTIC results specifically rather than the
lensing results collectively.  To be as fair as possible, we will
quote the range of lensing results as being 2.2--2.9$\sigma$
discrepant with the X-ray model.  For the dynamical model, the p-value
from the largest and most inclusive lensing dataset is 0.25 and in
Gaussian terms the discrepancies range from 0.7--1.4$\sigma$, which by
most standards is good agreement.

Third, even 2.2--2.9$\sigma$ is likely to be an overestimate of the
X-ray discrepancy because of the halo concentration value $c$ used by
C12.  A widely used approach in cluster modeling is to adopt the
concentration-mass relation found in N-body simulations such as in
Duffy {\it et al.}  (2008), whose Figure 2 shows that $c \approx 4$ is
the most reasonable estimate for a halo with the mass and redshift of
West.  C12 instead chose two illustrative values: $c=10$ for their
Table~4 and $c=5$ for their Table~5.  However, $c=10$ is highly
disfavored according to the Duffy {\it et al.}  (2008) results.  This
is why we have focused on C12's results for $c=5$ (their Table 5).
Furthermore, because the tension between the X-ray model and the
lensing data is substantially reduced as the model changes from $c=10$
to $c=5$, the tension is likely to be reduced even more for a halo
with $c=4$.  At the least, C12 should have considered a range of
concentrations extending below 5, and this would have reduced the
tension between the X-ray and lensing results.

In summary, the correct interpretation of the lensing analyses in C12
is that they are {\it at most} 2.2--2.9$\sigma$ low compared to the
X-ray model, and in good agreement with the dynamical model.  These
results are consistent with the rest of the literature.  C12 cite
Abate {\it et al} (2009) as finding a best-fit NFW mass of
$M_{200}=0.0^{+0.5}_{-0.0} \times 10^{14}$ as supporting evidence for
the low signal in ``A781D.'' As noted in \S\ref{sec-intro}, however,
Abate {\it et al} (2009) did {\it not} fit the West cluster; rather,
their ``d'' refers to S08's Subcluster which is an extension of the
gas in the Main cluster.  Once this mistake is rectified it is clear
that the only published weak lensing mass estimate of West is the one
by S08, which C12 overlooked and which is actually {\it
  higher} than the dynamical estimate.

We have demonstrated the consistency of weak lensing results with the
X-ray and dynamical results simply by considering the nonindependence
of data sets and by a careful reading of the literature.  Yet a new
question arises: why have different analyses of the same DLS lensing
data resulted in even $\sim 1\sigma$ variations in our view of
West?\footnote{For example, the C12 S/N is on the low side of
  estimates from the dynamical model whereas the S08 lensing model is
  on the high side of the dynamical model, and C12 were themselves
  motivated by convergence maps showing West as less prominent than
  the equally-massive East.}  Given that these analyses use the same
data set, one would expect much less variation unless the choice of
analysis procedure is the dominant source of variation; and in that
case, we should identify which choices in the analysis procedure are
so important.  We investigate this issue in the next section.


\section{Convergence map with source redshift weighting}\label{sec-willsmap}

The DLS has photometric redshifts (Schmidt \& Thorman 2013) which were
not available or not used for the convergence maps made by Wittman
{\it et al.} (2006), Khiabanian \& Dell’Antonio (2008), and Kubo {\it
  et al.}  (2009).  Since we know the cluster redshift, we can use the
source photometric redshifts to optimize sensitivity to lenses at this
redshift.  Photometric redshifts not only provide a much more highly
tailored cut against foreground objects compared to the magnitude and
size cuts used by C12, but they can also be used to lower the weight
of sources which are just behind the cluster and which are therefore
lensed inefficiently.  (Those sources are lensed more efficiently by
structures at lower redshift than the target cluster.)  This is
related to weak lensing tomography: in tomography we make maps from a
series of source redshift bins to explore changes in structure as a
function of redshift, whereas here we weight the source redshifts to
highlight structure at a known redshift.

Following Dawson {\it et al.} (2012), we note that the shear $\gamma$ is
proportional to a ratio of angular diameter distances:
\begin{displaymath}
\gamma \propto \frac{D_{ls}(z_l,z_s)D_l(z_l)}{D_s(z_s)}\mathcal{H}\left(\frac{z_s}{z_l}-1\right)
\end{displaymath}
where $D_l$, $D_s$, \& $D_{ls}$ are the angular diameter distances
from observer to lens, observer to source, and lens to source,
respectively, $z_l$ and $z_s$ are the lens and source redshifts
respectively, and $\mathcal{H}$ is the Heaviside step function.  A
matched filter for lenses at $z_l=0.43$ would then weight each source by
the distance ratio on the right-hand side with $z_l$ set to
0.43. However, for each source $z_s$ is a probability distribution
$p(z)$ rather than a single number.  We integrate over this
distribution to obtain the redshift-dependent part of the weight for
each galaxy:
\begin{displaymath}
w = \int
\frac{D_{ls}(z_l,z_s)D_l(z_l)}{D_s(z_s)}\mathcal{H}\left(\frac{z_s}{z_l}-1\right)
p(z_s) dz_s
\end{displaymath}
where $z_l$ is fixed at 0.43.  We also weight by the inverse variance
of the galaxy's shear measurement.

\begin{figure}
\centerline{\includegraphics[scale=0.3]{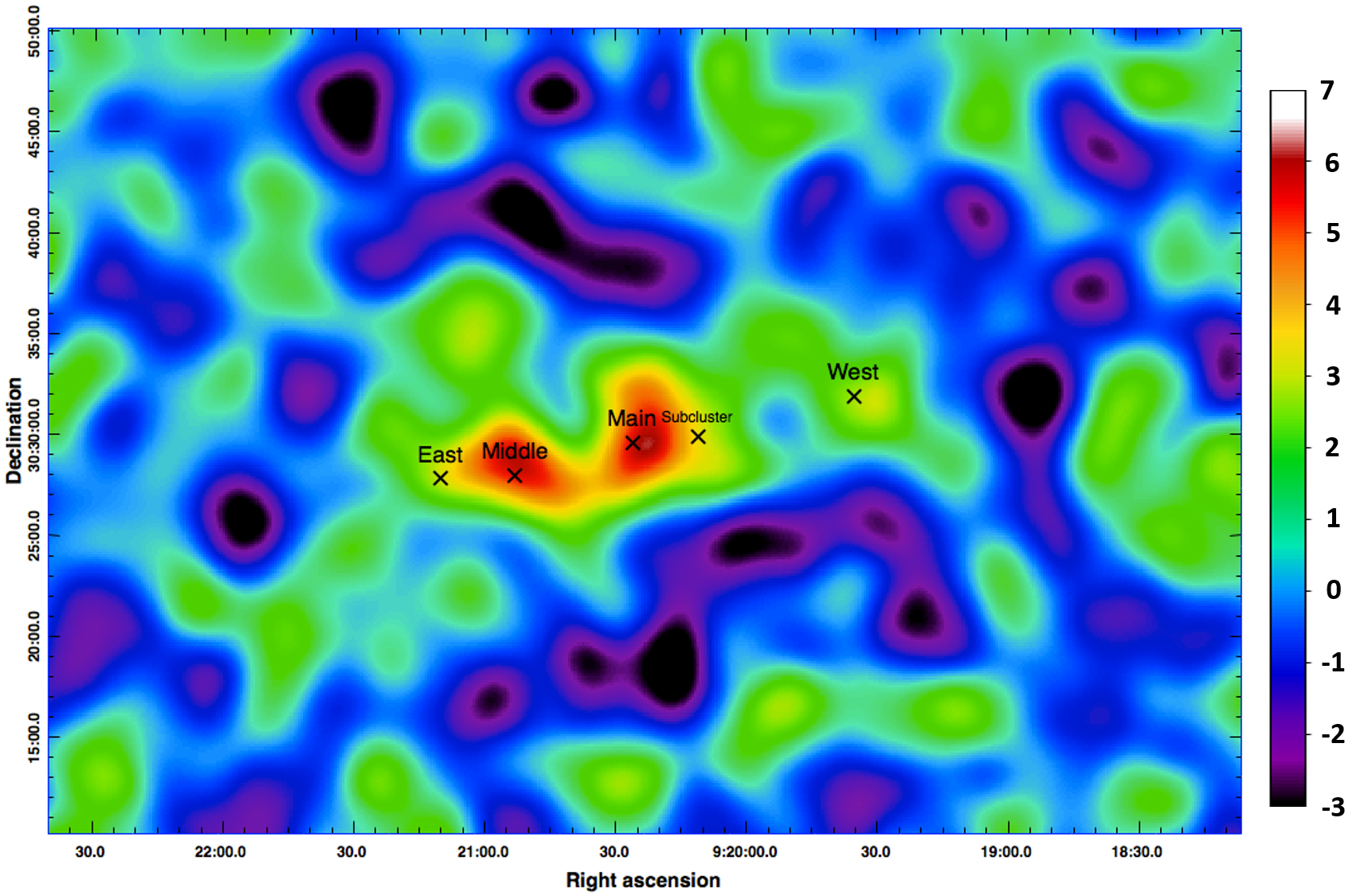}}
\centerline{\includegraphics[scale=0.3]{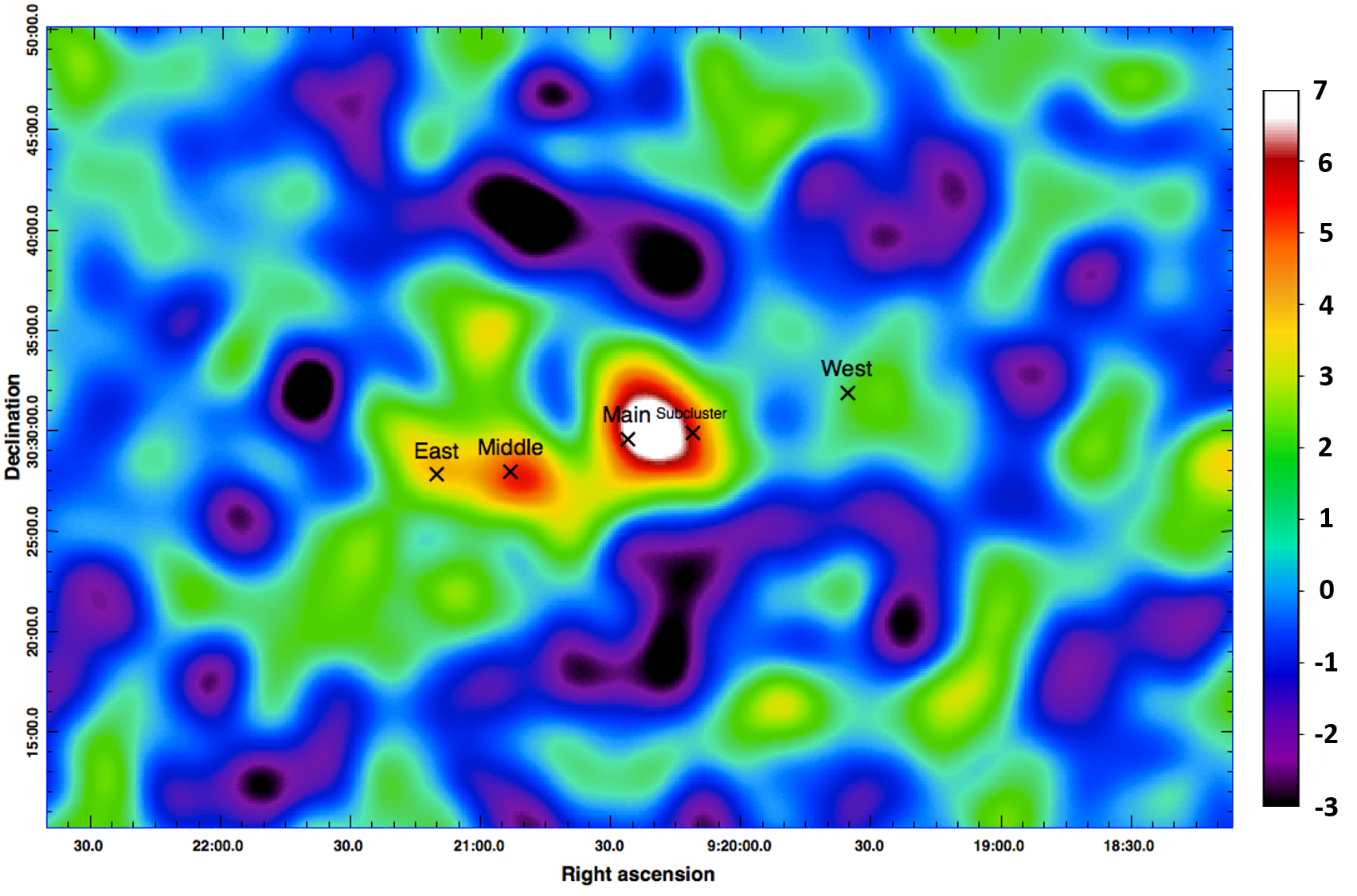}}
\caption{Top: Convergence map (in units of signal-to-noise, based on
  100 bootstrap realizations) of the Abell 781 region made with
  photometric redshift probability density ($p(z)$) weighting.  The
  West cluster is clearly visible at about the same signal-to-noise as
  the East cluster.  Bottom: the same map made with the C12 magnitude
  cuts rather than the $p(z)$ weighting.  Here, as in the convergence
  maps which motivated the C12 investigation, West appears at lower
  signal-to-noise than East.  The difference is consistent with
  shape-noise fluctuations (ie, chance alignments of source galaxies)
  at the low redshifts which are downweighted by the redshift
  weighting.
  \label{fig-willsmap}}
\end{figure}

The resulting convergence map of the Abell 781 region for $z_l=0.43$
is shown in Figure~\ref{fig-willsmap}, top.  In contrast to the maps
of Wittman {\it et al.} (2006), Khiabanian \& Dell’Antonio (2008), and
Kubo {\it et al.} (2009), the West cluster is about as prominent as
the East cluster, with $S/N=3.0$.  We attribute this to the source
redshift weighting.  Assigning different weights to each source
effectively resamples the data and results in a different realization
of shape noise.  Thus, this simple analysis choice can influence the
results about as much as the dominant source of noise because it
resamples the dominant source of noise.  To confirm this, we remade
the map with the C12 magnitude cuts rather than source redshift
weighting and indeed the S/N of West declines to 2.4 while that of
East increases slightly (Figure~\ref{fig-willsmap}, bottom).

Therefore, a plausible scenario which accounts for all the published
results is that the assumption that all sources share the same
effective distance ratio happens to bias the West signal low.  This
scenario does not require any specific source redshift distribution
near the West line of sight; it merely requires that the random
orientations of lower-redshift sources (including those behind the
West cluster but at low distance ratio) happen to reduce the measured
tangential shear.  Because the signal-to-noise of East and West on
these maps are already rather low, a mere 1--2$\sigma$ shape-noise
fluctuation along the West line of sight would be sufficient to nearly
remove West as a peak on the map.  This scenario would also explain
the fact that S08, who used photometric redshift weighting in their
weak lensing analysis, did not find an anomalously low weak lensing
mass for West.

Another potentially important difference between S08 and C12 is that
S08 employed a model-fitting approach, whereas C12 (and the map makers
cited by C12) employed a data-smoothing approach.  For completeness,
we also present the results of a model-fitting approach.  We fit four
simultaneous NFW models to the DLS weak lensing data with full $p(z)$
weighting and with centers fixed by the X-ray locations (other details
of the fitting procedure are as described in Dawson {\it et al.}
2012).  For all subclusters, the resulting masses
(Table~\ref{tab-massestimates}) agree (within $1\sigma$) with all
previous mass estimates.  Our mass estimates are on average slightly
higher than the weak lensing estimates of S08, which we attribute to
our use of the full $p(z)$ rather than the one-point redshift
estimates used by S08; Wittman (2009) showed that the one-point
estimates are biased tracers of the underlying $p(z)$.  In Appendix A
we show that the data-smoothing approach is subject to some types of
errors to which the model-fitting approach is immune.  However, the
two smoothed maps we present here demonstrate that the choice of a
data smoothing approach by itself does not result in the reduced
prominence of West.  The choice of redshift weighting is more
important.

Although the previously published maps served as motivation for the
C12 investigation of the S/N of West in multiple lensing data sets,
C12 do not themselves compare East and West because East is outside
the footprint of their new data.  Therefore, we have fully addressed
the two independent but related issues with a single hypothesis which
fits all the existing results.  A 1--2$\sigma$ low-redshift shape noise
fluctuation could (1) nearly remove West as a peak in previously
published convergence maps, but allow it to appear as a $3\sigma$ peak
when sources are properly weighted by distance ratio (as seen in this
work and in S08); and (2) explain why the weak lensing S/N modeling of
C12 is $\sim 1\sigma$ low compared to the dynamical model while S08
found a weak lensing mass slightly higher than the dynamical model.

\section{Summary}
\label{sec-summary}

C12 focus on ruling out PSF modeling errors as an explanation for the
weak lensing S/N they measure for the West cluster, and they succeed
in ruling this out by considering three independent data sets.
However, their conclusion that ``A strong discrepancy remains between
the [weak] lensing mass of A781D and the mass estimates from
spectroscopic and X-ray measurements'' is not at all justified after
correcting the following errors in interpreting the statistics and the
literature:
\begin{itemize}
\item their three lensing constraints are very far from independent so
  that the joint constraint found by C12, indicating a $\sim 5\sigma$
  deviation from the X-ray model, should actually be considered at
  most a 2.2-2.9$\sigma$ result.  The actual significance must be less
  than this because C12 overestimate the halo concentration parameter
  and do not account for the uncertainty in the halo concentration.

\item recognizing the nonindependence of the lensing constraints also
  removes any tension with the dynamical model.  The tension with the
  dynamical model is also removed {\it even if} the lensing data sets
  are independent and one simply uses the correct method for combining
  independent p-values.

\item C12 misread the Abate {\it et al.} (2009) weak lensing estimate
  of the Subcluster as applying to the West cluster.  In fact, the
  only previously published weak lensing mass estimate of the West
  cluster (S08) is actually {\it higher} than the dynamical estimate.
\end{itemize}

The S08 weak lensing mass estimate is on the high side of the
dynamical estimate whereas the modeling of C12 shows it to be on the
low side.  Although this difference is not statistically significant,
one would not expect even a 1$\sigma$ difference between two studies
using the same underlying (DLS) data set unless some systematic
analysis choice had a substantial influence. We identify this analysis
choice as the use of distance ratio weighting using source photometric
redshifts. A 1--2$\sigma$ shape-noise fluctuation at low redshift would
affect the C12 result and the previously published convergence maps
which motivated C12 (Wittman {\it et al.} 2006, Khiabanian \&
Dell'Antonio 2008, and Kubo {\it et al.} 2009), but would be nearly
invisible to the method of S08 and the weighted convergence map we
present here.

\section*{Acknowledgments} We thank Ian Dell'Antonio for kindly
providing data analysis files and useful comments.  Funding for the
Deep Lens Survey has been provided by Lucent Technologies Bell Labs,
UC Davis, and NSF grants AST 0441072 and AST 0134753.  Observations
were obtained at Cerro Tololo Inter-American Observatory and Kitt Peak
National Observatory. CTIO and KPNO are divisions of the National
Optical Astronomy Observatory (NOAO), which is operated by the
Association of Universities for Research in Astronomy, Inc., under
cooperative agreement with the National Science Foundation.

\appendix

\section{Interpreting weak lensing S/N maps}\label{sec-causes}

This Appendix addresses some issues which are potentially important in
the type of modeling conducted by C12 but which turned out to be
subdominant in resolving the mystery of Abell 781 West.  Most
discussions of weak lensing errors, including that of C12, focus on
PSF modeling and shape measurement, but there is much more to
consider:

\begin{itemize}

\item{unmodeled foreground source contamination:} C12 use COSMOS (Scoville
  {\it et al.} 2007) data to model their source redshift distribution,
  without explicitly accounting for the presence of a rich cluster
  (Main) in the foreground of West.

\item{effect of nearby structures:} East may appear more prominent
  than West on a smoothed convergence map because shear from Middle is
  smoothed into East (to a far greater extent than shear from Main is
  smoothed into West).

\item{spatially varying noise:} the bright star projected near West
  causes a loss of source galaxies and therefore an increase in noise
  in the region around West.  In the data-smoothing approach, missing
  data in high-signal regions also causes a loss of signal.  The
  simulation approach adopted by C12 is capable of including these
  effects, but there is no evidence that C12 included them.

\item{map pixelization:} C12 used large (1.5\arcm) pixels and West's
  X-ray peak is near a pixel corner, so we examine the effect of a
  $\sim 1$\arcm\ miscentering.

\end{itemize}

Although many of these differences turned out to have little effect on
this particular result, we describe them briefly here in the hope that
students of weak lensing will benefit from the discussion.  In several
places, we point out advantages of a model-fitting approach over a
data-smoothing approach.

\subsection{Unmodeled foreground source contamination}


C12 imposed more or less standard size and magnitude cuts on their
sources.  Recognizing that these cuts would not eliminate all
foreground and cluster member galaxies, they modeled the contamination
by imposing the same cuts on the COSMOS catalog, ray-tracing the
resulting source catalog through a model lens, and comparing the
simulated S/N maps to their measured S/N maps.  But the presence of
Main in West's foreground implies the presence of more foreground
galaxies than in a random field. If these galaxies were present in the
final source catalog they would dilute the lensing signal.  To test
for this, we examined the photometric redshift distribution of sources
surviving their cuts in random DLS regions versus in the area around
West.  We did not find that the area around West contained an excess
of sources at the redshift of Main, after the C12 cuts.  This implies
that the maximum source size imposed by C12 was effective in
eliminating Main galaxies.  Therefore Main's contribution to
foreground contamination, although potentially an important unmodeled
effect, appears to have played little role in practice.

\subsection{Effect of nearby structures}

The data-smoothing approach does not allow us to distinguish between
shear from different sources.  Middle is projected only 3\arcm\ from
East so shear from Middle is smoothed into East's area.  Shear from
Main must also be smoothed into West's area, but given the larger
separation (11\arcm) this should be a much smaller effect.  For some
choice of smoothing scales, this could explain why East appears more
prominent on convergence maps but would not explain why the shear
observed in the direction of West falls short of model predictions.

We tested this hypothesis by generating mock source catalogs with
shear representing different cluster geometries (single cluster, two
clusters with the East-Middle separation, or two clusters with the
Main-West separation), and creating convergence maps with the same
{\it fiatmap} code and smoothing scales used by C12.  We did not find
a substantial boost for the mock East cluster above that of West; in
fact the boost in either case was minimal.  Therefore this effect does
not answer the question raised by C12, but readers should be aware of
the possibility when interpreting smoothed maps.  When considering
multiple lenses projected near each other, a simultaneous
model-fitting approach (S08, Abate {\it et al.} 2009) facilitates a
better distinction between the contributions of different lenses.

In general, weak lensing analyses should remain cognizant of the
uncertainty due to less easily identifiable foreground and background
structures, which can be tens of percent (Rasia {\it et al.} 2012 and
references therein).  C12 did recognize this and were able to discount
it using the redshift survey of Geller {\it et al.}  (2010) in this
field.

\subsection{Spatially varying noise}

Because C12's primary claim is that the S/N of West is low, it is
important to examine not only effects on signal but also sources of
noise.  Bright stars increase noise locally by masking background
galaxies, and the West cluster has a bright ($V=11.9$) star only $\sim
54$\arcs\ away.  The size of the effect depends on the details of the
size of the masked area, its geometry relative to the lens center and
other structures in the region, and (for the data-smoothing approach)
the filter used to smooth the data.  A loss of data will increase
uncertainty for both model-fitting and data-smoothing approaches, but
the data-smoothing approach is uniquely sensitive to a loss of signal:
if the unobserved area falls near the lens center, the smoothed shear
field will be biased low.  In contrast, unobserved area is simply
ignored when fitting a model, so that no bias can be introduced.
(However, note that VanderPlas {\it et al.}  2012 have proposed a
method for interpolating shear over masked areas.)  The sizes of these
effects depend on the geometry, and simulations are the best tool for
assessing the impact.

We simulated the effect of the lost area for the given geometry, using
a 1\arcm\ radius mask around the star. (C12 did not explicitly use a
mask, but scattered light from the star does remove some area, and we
chose a large mask as a starting point.) We created mock catalogs with
shear imprinted by four singular isothermal sphere lenses arranged in
the geometry of the Abell 781 region and created convergence maps with
the same {\it fiatmap} code used by C12, with and without placing a
mask near the West cluster.  We did not find a significant effect on
the S/N of the mock West cluster for this set of parameters.  A
possible contributing factor is that the Main cluster and the mask are
on opposite sides of the the West cluster.  This means that when
tangentially averaging around the West cluster, galaxies which are
more sheared by the Main cluster are retained while galaxies which are
less sheared by the Main cluster are masked.  Thus two weaknesses of
the data-smoothing approach---sensitivity to neighboring lenses and to
missing data---mostly cancel each other here.

\subsection{Convergence map pixelization}

A pixelized map represents some function of the shear field {\it
  evaluated at the center of each of its pixels}.  If the peak of the
lens happens to be near the corner of a pixel, this miscentering will
result in a smaller pixel value than if the peak of the lens happens
to be at the center of a pixel.  Normally this is not a large effect
because one can choose to make the pixel size as small as desired.
Small pixels do not improve the angular resolution of the map beyond
what is supported by the smoothing scale, but they do prevent
underestimation of the S/N due to miscentering.  C12 used large
(1.5\arcm) pixels and the X-ray position of West is near the corner of
one pixel, so the miscentering hypothesis must be investigated.

We again used mock catalogs to estimate the size of this effect for
the particular geometry here and for a singular isothermal sphere
(SIS) model as an extreme example (less peaky profiles produce a
smaller miscentering effect).  C12 provided us with a variety of maps
with a range of smoothing parameters.  We calculated the reduction in
signal due to miscentering for each choice of smoothing parameters,
and found that it ranged from 1--30\%.  In other words, the measured
signal could be as small as 70\% of the expected signal due to the
miscentering effect, but only if the true shear profile is as peaky as
an SIS.  The miscentering effect could therefore play a small role in
suppressing the S/N measured by C12.  However, if C12's mock
observations correctly incorporated West's X-ray peak location
relative to their map grid, then this effect is already included in
their S/N modeling results and therefore would not alter their
results.

\end{document}